# Harnessing near-field thermal photons with efficient photovoltaic conversion

C. Lucchesi[1], D. Cakiroglu[2], J.-P. Perez[2], T. Taliercio[2], E. Tournié[2], P.-O. Chapuis[1], R. Vaillon[2,1]

[1]Univ Lyon, CNRS, INSA-Lyon, Université Claude Bernard Lyon 1, CETHIL UMR5008, F-69621, Villeurbanne, France

[2]IES, Univ Montpellier, CNRS, Montpellier, France

**A huge amount of thermal energy is available close to material surfaces in radiative and non-radiative states[1–3], which can be useful for matter characterization[4–6] or for energy devices[7,8]. One way to harness this near-field energy is to scatter it to the far field[4–6]. Another way is to bring absorbers close to thermal emitters[9,10,2,3], and the advent of a full class of novel photonic devices exploiting thermal photons in the near field has been predicted in the last two decades[7,11,12,13,14]. However, efficient heat-to-electricity conversion of near-field thermal photons, *i.e.* the seminal building block, could not be achieved experimentally until now. Here, by approaching a micron-sized infrared photovoltaic cell at nanometric distances from a hot surface, we demonstrate conversion efficiency up to 14% leading to unprecedented electrical power density output (7500 W.m$^{-2}$), orders of magnitude larger than all previous attempts. This proof of principle is achieved by using hot graphite microsphere emitters (~800 K) and indium antimonide cells, whose low bandgap energy matches the emitter infrared spectrum and which are specially designed for the near field[15]. These results pave the way for efficient photoelectric detectors converting thermal photons directly in the near field. They also highlight that near-field thermophotovoltaic converters[16], which harvest radiative thermal energy in a contactless manner, are now competing with other energy-harvesting devices, such as thermoelectrics, over a large range of heat source temperatures[17].**

A significant number of experimental demonstrations of the enhancement of thermal radiation heat transfer between a hot and a cold body[18–22] establish a firm basis for new photonic devices based on thermal photons in the near field, involving for instance thermal rectification[13] and photonic cooling[14]. In particular, it was predicted 20 years ago that near-field effects would drastically increase thermal radiation transfer between a hot emitter and a photovoltaic cell, and in turn electrical power generation[23–25]. Many theoretical results have further elaborated on this idea[8,26–30]. An early experiment[31] showed some qualitative enhancement of the photogeneration of electrical charges at microscale gaps between the emitter and the cell. More recently, three experimental works[32–34] reported on an enhancement of the electrical power in the near field, with estimations of efficiency (when available) of 0.02% and 0.98% and a maximal power density of 7.5 W.m$^{-2}$. These modest performances indicate that a clear proof of efficient photovoltaic conversion of near-field thermal radiation, leading to both significant output power and conversion efficiency, is still lacking. This is puzzling since thermophotovoltaics, based on the conversion of radiative heat emitted in the far field, has recently been demonstrated to be ultra-efficient[35]. Nevertheless, the electrical power density measured for these far-field devices remains moderate (approaching the W.cm$^{-2}$ level[35,36] only for high-temperature emitters above 1000°C) in comparison to other thermal energy-harvesting devices[37]. Near-field thermophotovoltaic converters, where the radiative power transferred is much higher, have the potential to mitigate this issue[17].

In the present work, we have specifically designed and fabricated an indium antimonide (InSb) cell having a very low bandgap energy (0.23 eV, *i.e.* 5.3 μm, at 77 K), which is well suited for conversion of thermal radiation emitted from medium-grade heat sources (temperatures in the range 250-650°C) and therefore resolves one of the obstacles to enhanced efficiency. We have also identified cell design parameters that allow the maximization of radiative heat-to-electricity conversion in the near field. In addition, we have chosen the configuration of a spherical emitter exchanging radiation with a planar receiver[18,19,38–41], which both improves the view factor between the emitter and the cell and allows probing a large emitter-cell distance range, from millimeters down to nanometers.

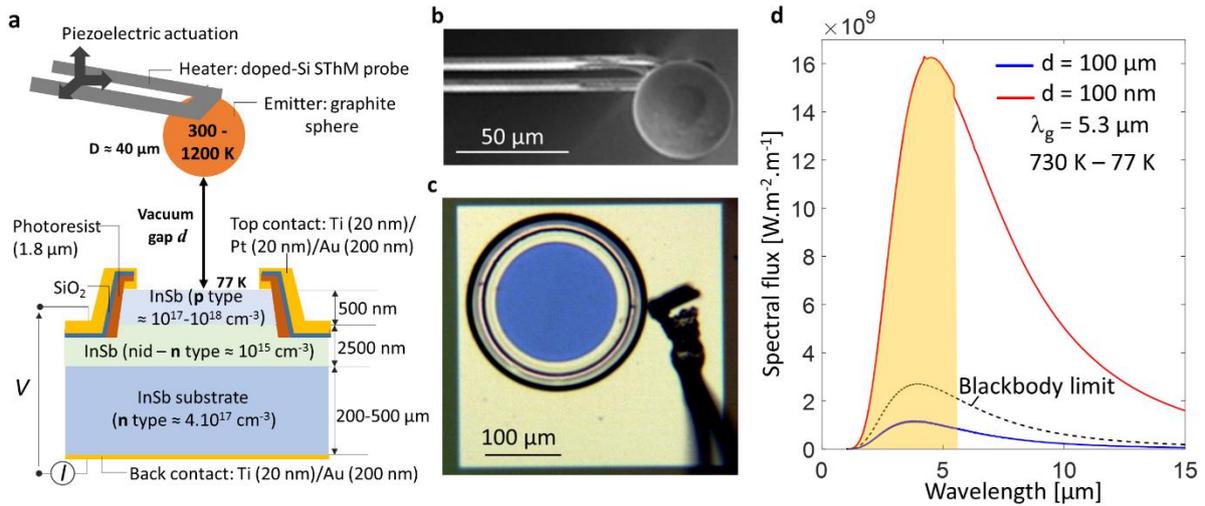

**Figure 1 | Device and simulated radiative heat transfer for photovoltaic conversion of near-field thermal photons.** a) Scheme of the emitter-photovoltaic cell device. b) Scanning electron micrograph of the emitter seen from below. A graphite sphere is glued on the tip of a SThM doped-Si probe with a ceramic adhesive. c) Optical microscopy top view of an InSb cell. d) Simulated thermal radiation heat transfer spectrum for the case of a 732 K emitter and 77 K cell for far-field (blue curve) and near-field (red curve) configurations.

The experimental set-up (see Fig. 1a) involves a micron-sized photovoltaic cell made of a thin InSb p-on-n junction (view from top in Fig. 1c) and a spherical graphite emitter glued to the probe of a scanning thermal microscope (SThM, see Fig. 1b). A key feature is that the infrared photovoltaic cell is specifically designed for harvesting thermal radiation fluxes of medium-grade temperature sources, around 500°C, in the near field[42]. It is made of InSb, one of the III-V semiconductors with the longest bandgap wavelength, to be able to convert photons of wavelengths in the micrometer range (see Fig. 1d). The bandgap wavelength depends on temperature and varies from 5.3 μm at 77 K to 7.3 μm at room temperature. The architecture of the cell was optimized based on full calculations involving the coupling of charge transport and radiative absorption in a 1D configuration[42]. The junction is located close to the cell top surface (at few hundreds of nanometers only) so as to collect a large share of the near-field photons. In order to avoid having zones of the cells which are not illuminated (generating a dark current without any photocurrent), the size of the cells is matched to that of the emitter, which increases the view factor. Thus the cell geometries involve mesas with active-area diameters of the order of few tens of micrometers (see Methods and Ext. Data Fig. 7a). Gold layers are used as top and bottom electrical contacts. It is important to underline that InSb can operate as a proper p-n junction only at low temperature[15], so in this experiment the cell is cooled down to 77 K by placing it on the cold finger of a cryostat located in a vacuum chamber (see Methods).

The emitted radiative heat flux is measured by means of a scanning thermal microscope, which is an atomic force microscope (AFM) where a temperature sensor is located on the cantilever[43]. A resistive sensor made of doped silicon is used, the evolution of which as a function of temperature is known. The temperature of the sphere is estimated to be equal to that of the resistor and the thermal conductance from the tip temperature to the ambient was measured, which allowed monitoring variations of the radiative flux lost by the emitter (see Methods and Ext. Data Fig. 1). Most of the results presented here were obtained for a tip apex temperature of 732 K (439 °C), which maximizes measurement sensitivity, but the emitter could be heated up to temperatures larger than 1200 K. Graphite was selected for the emitter as the material maximizing the near-field radiative exchange with InSb (see Supplementary Information (SI) and Ext. Data Fig. 2). Fig. 1d indicates that about 40% of the overall power exchanged between a planar emitter and a planar receiver is located in the wavelength range that can be converted to electricity. It also shows that reducing the distance between the emitter and the cell strongly increases radiative exchange. In the experiment, a sphere is therefore approached in the vicinity of the cell by moving it vertically with a piezo-actuator. A spherical emitter was selected, despite its reduced near-field exchange with respect to a planar one at the same distance (see Ext. Data Fig. 2c and SI), in order to eliminate the parallelism issues inherent to planar configurations. Such geometry was successful for Casimir force and near-field radiative heat flux experiments[18,19,38–41,43] and allows probing a large emitter-cell distance range, from the sub-100 nm region to millimeters.

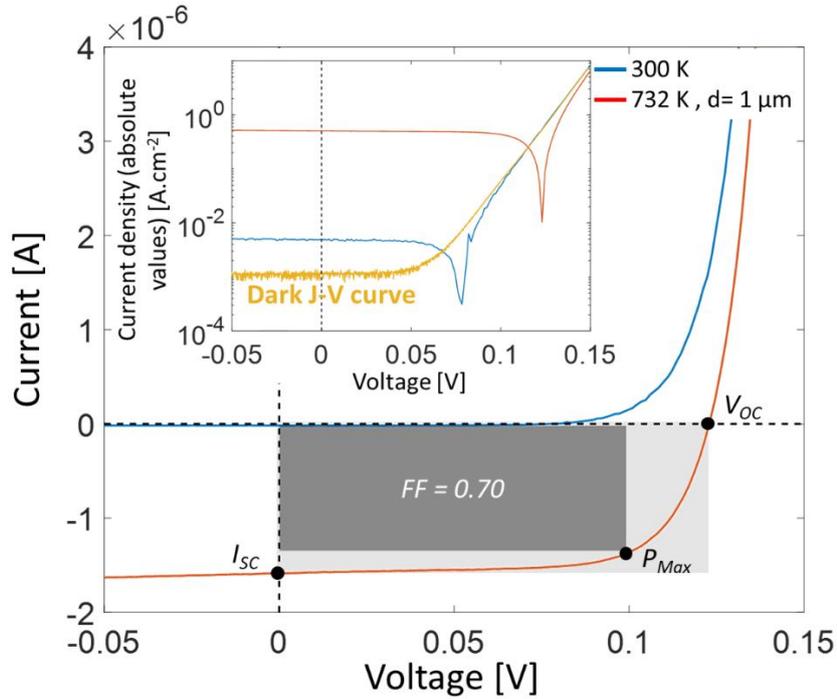

**Figure 2 | Current-voltage characteristics of a cell at 77 K having a 20 µm active area diameter.** I-V curves under 300 K ambient illumination (blue), 732 K graphite sphere at *d* = 1 µm (red) and under dark conditions (orange, inset). The dark and light gray rectangles represent respectively the maximum output power configuration and the product $V_{OC} I_{SC}$, used for computation of the fill factor (*FF*). The inset shows the current density in absolute values according to the applied voltage. The curves were obtained for a 500 µm-thick InSb substrate and a $10^{17}$ cm$^{-3}$ p-doping level.

The cells are first characterized at a fixed distance from the emitter (see Fig. 2). I-V curves in the dark indicate a low noise level (mA.cm$^{-2}$) in the experiments. It is remarkable that room-temperature radiative emission leads to a significant electrical power generation by the cell. The open-circuit voltage depends on the temperature of the emission source (see diverging values in the logarithmic scale of the inset of Fig. 2) and is found to be larger than 100 mV at 77 K, which is around half the bandgap energy of InSb ($E_g$ = 0.23 eV). In contrast to previous reports[33,34], the I-V curves look like usual photovoltaic cell characteristics, i.e. rectangles with a rounded corner. The short-circuit currents depend on the emitter temperature and reach already microamps at a distance of 1 µm. Fill factors as high as 0.75 were found, leading to electrical output powers up to 5.5 µW, depending on the distance.

The high sensitivity of the electrical power output measurement allows following the approach of the hot sphere towards the cold cell from distances up to millimeters. It is observed that variations of the short-circuit current as a function of the emitter-sphere distance superimpose nicely the prediction of far-field view factor theory for these large distances (see inset in Fig.

3b and Ext. Data Fig. 8b), indicating that far-field thermophotovoltaic conversion depends only on the level of illumination.

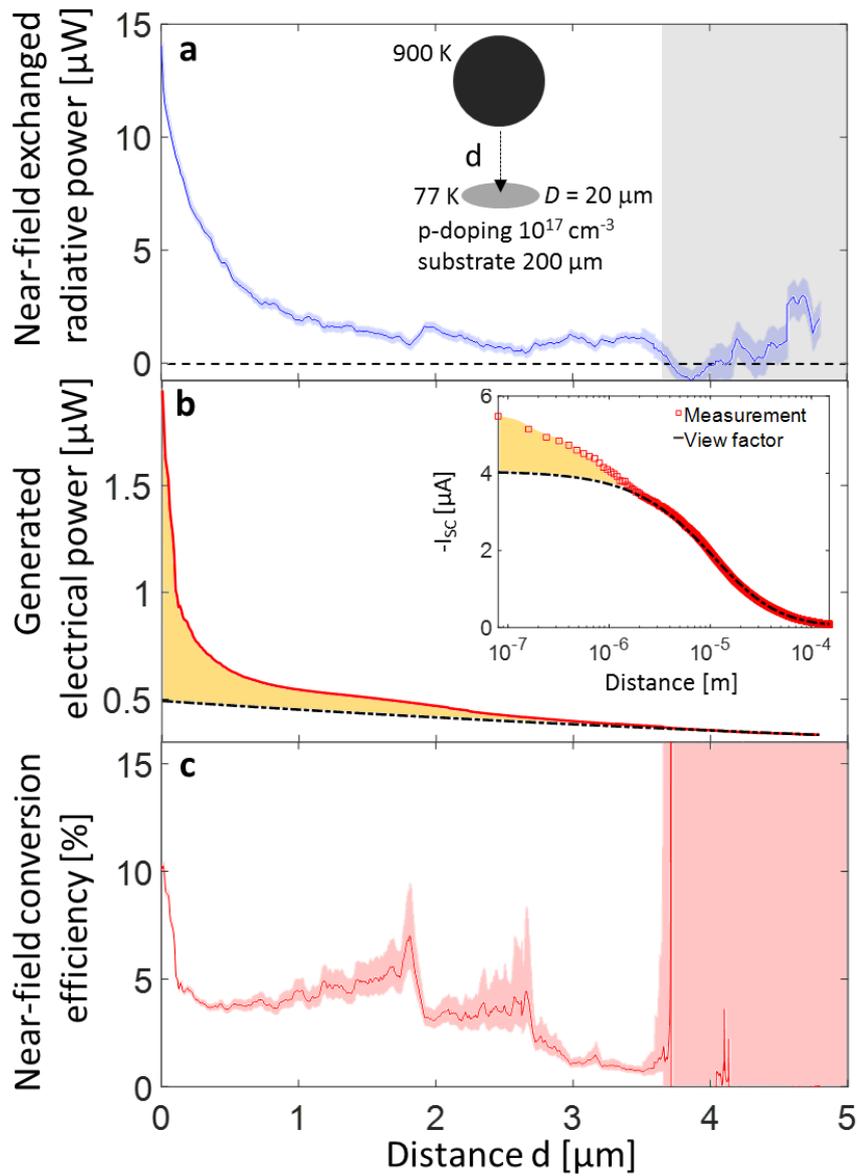

**Figure 3 | Near-field exchanged radiative power, maximum generated electrical power and near-field conversion efficiency obtained simultaneously as a function of the emitter-cell distance.** a) Near-field radiative power according to distance. b) Generated electrical power, deduced from the measurement of the cell short-circuit current and the application of the superposition principle, as a function of distance. The total electrical power (red curve) is compared to the power increase due to the far-field contribution estimated from the view factor (dashed-dotted black curve). The region between the two curves provides the near-field electrical power. The inset data, obtained with another cell, show the match between the view factor computation (plain line) and the experimental data in the far field. c) Near-field conversion efficiency calculated by dividing the near-field electrical power (b) by the near-field exchanged radiative power (a).

We now focus on the near field, which is approximately given by the distance range below Wien's wavelength ($\lambda_W.T$= 2898 µm.K) for the emitter, *i.e.* below 4 µm. Our setup allows the measurement of the radiative exchange as a function of the emitter-cell distance and the increase of the short-circuit current at the same time. As expected, a strong enhancement of the flux is observed for small distances, where the exchanged radiative power exceeds 14 µW. Almost all this increase is due to the near-field contribution, since the gain from the far-field contribution is much weaker in the last micrometers. While we cannot directly measure the far-field contribution background, it is computed to be ~7.8 µW for a sphere at 900 K and a cell having a 20 µm active area diameter (see theoretical analysis in SI and Ext. Data Figs. 2-3).

The electrical output power can be derived from the short-circuit current data obtained as a function of emitter-cell distance at the same time as the radiative flux data. We show in Ext. Data Fig. 9 that all I-V curves can be superimposed onto the dark I-V curve by shifting the short-circuit current to zero ("superposition principle"). Hence by measuring only the short-circuit current as a function of distance, one can recover all illuminated I-V curves and deduce the maximum electrical power output as a function of distance (Fig. 3b, red curve). It reaches almost 2 µW for the selected case, leading to an overall efficiency estimated to be about 9%. This is one order of magnitude higher than the best previous experiment[32]. It is noticeable that such a value competes with the efficiency of state-of-the-art thermoelectrics[44,45]. It is interesting to estimate the contribution of the near field to this power increase. By matching the view factor computation as a function of distance to the largest distances in the curve (dashed-dotted black curve in Fig. 3b), we are able to single out a stronger increase of the electrical output power in the sub-3 µm region than what standard far-field theory predicts (yellow region). Electrical output power 4.8 times larger than the far-field one is found.

The photovoltaic conversion efficiency of the near-field flux can be determined from our experimental data as a function of emitter-cell distance in the last 3 µm. By dividing the electrical power by the near-field radiation power, we find a near-field conversion efficiency of the order of 5 % for distances between 0.2 and 2.7 µm (uncertainty is linked to the near-field radiation data), with a strong increase for the closest distances to a value above 10%. The distance is measured by the vertical piezoelectric actuator, with an uncertainty close to the contact up to around 80 nm (see Ext. Data Fig. 6 and SI).

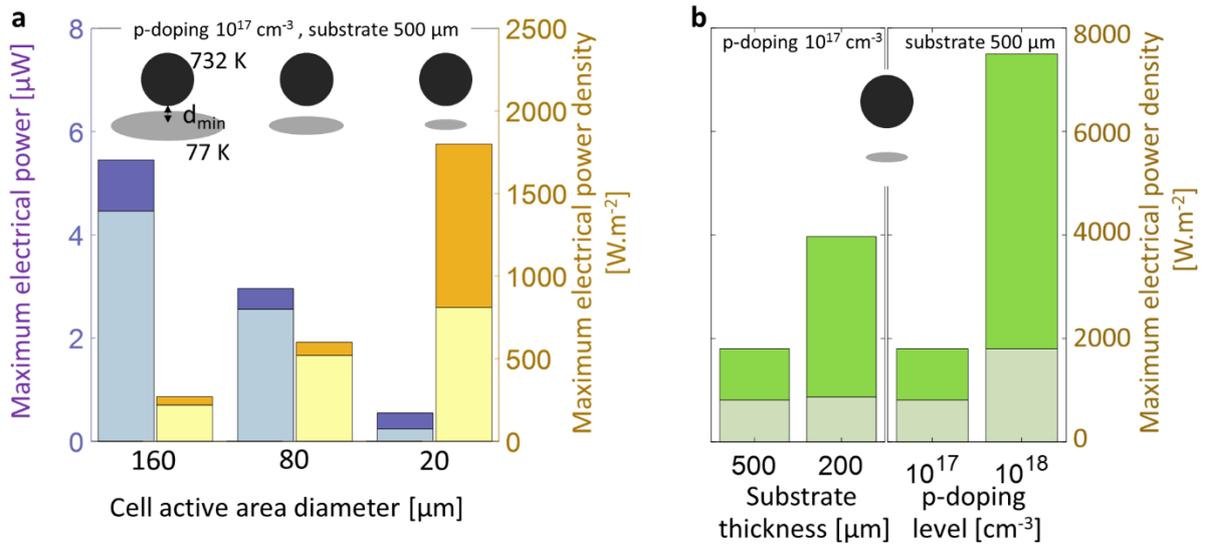

**Figure 4 | Influence of cell diameter, substrate thickness and p-doping level on the maximum electrical power for the smallest emitter-cell distances.** The dark and light parts are respectively the far-field and near-field contributions. a) Influence of the cell active area diameter. The electrical power decreases with the cell diameter because less illumination is collected, but the electrical power density increases. b) (left) Influence of the substrate thickness. The maximum electrical power density is observed for the device with the thinnest substrate. (right) Influence of the p-doping level. The maximum power density increases with the doping level.

Parameters that can enhance radiative heat-to-electricity conversion are now analyzed. First, the effect of the diameter of the active area of the cell is reported in Fig. 4a. As a larger share of the exchanged power comes from the near field for the smallest cell (dark orange region), it performs the best, being able to generate about 2 kW.m$^{-2}$, which is three orders of magnitude higher than previous reports[32–34]. Due to the high mobility of electrons in InSb, critical electrical losses are unlikely. However, parasitic absorption losses are expected for photon energies above the bandgap. Note that the near-field contribution in the sphere-plane geometry is mostly related to optical modes that are evanescent in the gap but propagate in the substrate ("frustrated modes", see SI and Ext. Data Figs. 4-5). By reducing the substrate thickness from 500 µm to 200 µm, part of the parasitic absorption is mitigated and a second passage of these photons across the junction is possible after reflection by the bottom gold contact layer which acts as a mirror[46]. A clear enhancement is observed in the experiment (Fig. 4b (left)). In addition, the doping level of the p-region of the junction was also shown to be a key parameter for the cell performances[42]. An impressive 4-fold enhancement of the electrical power is observed when increasing the dopant concentration from 10$^{17}$ to 10$^{18}$ cm$^{-3}$, leading to a record 7.5 kW.m$^{-2}$ value for the tested micron-sized cells, despite the moderate temperature of the emitter (432°C). Let us analyze the results in terms of efficiency. Near-field efficiency for the cell with a 500 µm-thick substrate and a $10^{17}$ cm$^{-3}$ p-doping concentration was directly measured to be 5.8% (lower than that reported in Fig. 3 due to lower temperature in Fig. 4). By thinning down the substrate of the cell, we find that the near-field efficiency, also measured directly from the output power, reaches 14.1 % (see also Ext. Data Tab. 1 and SI). This high value is explained first by the increase in power conversion for photon energies above the bandgap and second by the reflection of near-field thermal photons below the bandgap, which reduces the parasitic absorption that does not lead to photoconversion[42,47].

In summary, we have demonstrated that near-field thermal energy can be exploited efficiently. We have measured photovoltaic conversion of thermal photons in the near field with efficiency beyond 10% and electrical power densities reaching 7.5 kW.m$^{-2}$, close to the ~1 W.cm$^{-2}$ threshold typical for powerful energy harvesting devices[48] and close to the best far-field thermophotovoltaic performances despite much lower emitter temperatures[35,36]. This proof of principle for efficient near-field thermophotovoltaic conversion paves the way for fabrication of devices in different fields. Scattering near-field energy to the far field allows collecting only a weak fraction of it, which may lead to signal-to-noise ratio issues in thermal nanocharacterization setups[4–6]. The present approach highlights the possibility of using the photocurrent as signal, converting the photons directly close to the surface with a superior sensitivity. For energy harvesting as in solar thermophotovoltaics[49–51], it is also required to upscale the sizes and to avoid cooling the cell. Our architectures are compatible with more industrial-size implementations involving parallel flat surfaces that were undertaken recently[52]. Recent strategies for designing efficient cells with narrow energy bandgap III-V materials operating at room temperature[53] can also be applied for frustrated photon modes and will be particularly useful for harnessing energy of medium-grade heat sources. For high-grade heat sources, we have demonstrated the possibility to maintain a temperature difference larger than 1100 K across a small distance (see Ext. Data Fig. 1 and SI Sec. 10), so larger energy gap materials used in cells operating at room temperature could also be considered with a careful design for the near field following the highlighted paths. Finally, we remind that further enhancement of the performances is predicted for materials with polariton resonances above the energy bandgap[26,27].

**Main references**

1. Shchegrov, A. V, Joulain, K., Carminati, R. & Greffet, J.-J. Near-Field Spectral Effects due to Electromagnetic Surface Excitations. *Phys. Rev. Lett.* **85**, 1548–1551 (2000).

2. Joulain, K., Mulet, J.-P., Marquier, F., Carminati, R. & Greffet, J.-J. Surface electromagnetic waves thermally excited: Radiative heat transfer, coherence properties and Casimir forces revisited in the near field. *Surf. Sci. Rep.* **57**, 59–112 (2005).

3. Volokitin, A. I. & Persson, B. N. J. Near-field radiative heat transfer and noncontact friction. *Rev. Mod. Phys.* **79**, 1291–1329 (2007).

4. De Wilde, Y. *et al.* Thermal radiation scanning tunnelling microscopy. *Nature* **444**, 740–3 (2006).

5. Jones, A. C. & Raschke, M. B. Thermal infrared near-field spectroscopy. *Nano Lett.* **12**, 1475–81 (2012).

6. Weng, Q. *et al.* Imaging of nonlocal hot-electron energy dissipation via shot noise. *Science* **360**, 775–778 (2018).

7. Tervo, E., Bagherisereshki, E. & Zhang, Z. Near-field radiative thermoelectric energy converters: a review. *Front. Energy* **12**, 5–21 (2018).

8. St-Gelais, R., Bhatt, G. R., Zhu, L., Fan, S. & Lipson, M. Hot Carrier-Based Near-Field Thermophotovoltaic Energy Conversion. *ACS Nano* **11**, 3001–3009 (2017).

9. Polder, D. & Van Hove, M. Theory of radiative heat transfer between closely spaced bodies. *Phys. Rev. B* **4**, 3303–3314 (1971).

10. Pendry, J. B. Radiative exchange of heat between nanostructures. *J. Phys. Condens. Matter* **11**, 6621 (1999).

11. Liu, X., Wang, L. & Zhang, Z. M. Near-field thermal radiation: Recent progress and outlook. *Nanoscale Microscale Thermophys. Eng.* **19**, 98 (2015).

12. Ben-Abdallah, P. & Biehs, S. A. Thermotronics: Towards Nanocircuits to Manage Radiative Heat Flux. *Zeitschrift fur Naturforsch. - Sect. A J. Phys. Sci.* **72**, 151–162 (2017).

13. Fiorino, A. *et al.* A Thermal Diode Based on Nanoscale Thermal Radiation. *ACS Nano* **12**, 5774–5779 (2018).

14. Zhu, L. *et al.* Near-field photonic cooling through control of the chemical potential of photons. *Nature* **566**, 239–244 (2019).

15. Cakiroglu, D. *et al.* Indium antimonide photovoltaic cells for near-field thermophotovoltaics. *Sol. Energy Mater. Sol. Cells* **203**, 110190 (2019).

16. Basu, S., Zhang, Z. & Fu, C. Review of near-field thermal radiation and its application
10

**Methods**

**Fabrication of the emitter**

The emitter (Fig. 1b) consists of a commercial doped-Si scanning thermal microscopy probe (VITA-HE-NANOTA-200 from Bruker) on which a graphite sphere was glued using a 600 nm particle size $Al_2O_3$-based ceramic adhesive (RESBOND 989F) which can withstand temperatures up to 1920 K. The sphere of diameter 37.5 µm (Carbon powder from Goodfellow) was heated by the doped-Si tip of the SThM probe. Graphite was selected from an analysis of various materials (see Ext. Data. Fig. 2a,b). The electrical resistance of the probe and its dependence on temperature allowed simultaneous temperature measurements and heating by Joule effect of the sphere up to approximately 1200 K. Above this temperature and until the melting temperature of silicon (1687 K), the electrical resistance variations are insufficient to perform accurate temperature measurements, but the emitter can still be heated (see Ext. Data Fig. 1a).

**Calibration curve of the emitter temperature**

Each emitter had to be calibrated in temperature by measuring the electrical resistance dependence on temperature of the scanning thermal microscopy probe to which it was glued. First, the whole system composed of the SThM probe and its half-moon shaped holder was put in an oven typically used for thermocouples calibration (Fluke 9144). Because of the holder, the electrical resistance measurement could only be performed from room temperature to 413 K (140 °C). At higher temperatures, the glue and other components of the holder started to deteriorate and the probe could be damaged. To retrieve the $R(T)$ curve at these higher temperatures, a measurement of the resistance according to electrical power was performed. A current source provided an electrical current to the probe up to 11 mA while the probe voltage was measured so the resistance could be calculated. This kind of doped-Si probe has a characteristic $R(T)$ curve with a maximum resistance $R_{max}$ for a given temperature $T_{Rmax}$ which depends on the doping level of the low-doped part of the probe, located near the tip (Ext. Data Fig. 1). At this point, the thermally generated carrier concentration becomes higher than the doping level and the temperature evolution becomes linear with electrical power[54]. For the resistance measurement at $T < T_{Rmax}$ made in the oven, the data could be well fitted with a quadratic model up to a few kelvins before $T_{Rmax}$. For $T > T_{Rmax}$ the temperature was calculated from the electrical power by multiplying with a fitting factor so both temperature scales matched near $T_{Rmax}$. Put together, the data from the direct measurement in the oven and the $R(P)$ electrical measurement gave the $R(T)$ curve for the entire temperature range (see Ext.. Data Fig. 1). The temperature coefficient of the electrical resistivity (TCR) α in K$^{-1}$ can be calculated as $\alpha = \frac{1}{R}\frac{dR}{dT}$ and is useful to determine the temperature $T_S$ for which the sensitivity is maximum. For these SThM probes, $T_S$ is around 732 K and corresponds to the maximum value (in absolute values) of α, where a small change in temperature induces a large change in electrical resistance that can be measured more easily. It can also be noticed that beyond around 1200 K the sensitivity becomes very low and it is therefore complicated to accurately determine the temperature based on the resistance measurement.

**Temperature measurement of the emitter and near-field thermal conductance**

Measurement of the emitter temperature $T$ is needed to determine the radiative heat transfer with the cell characterized by the near-field thermal conductance $G_{NF}$. The principle is to measure the electrical resistance of the SThM probe $R$, thermalized with the emitter, and use calibration data to infer its temperature (see above). A Keithley 6221 signal generator was used to provide DC current $I$ to the SThM probe, up to 8 mA depending on the required temperature. A Wheatstone bridge circuit with a 100x voltage amplification was used to accurately measure the electrical resistance. The bridge was balanced with the emitter out of contact, for a distance $d$ slightly smaller than 5 µm, by adjusting the current supply until the electrical resistance of the probe matches the one imposed by a programmable resistance (MEATEST M632) corresponding to the targeted temperature. During the approach, the imbalance of the bridge was measured according to the displacement of the emitter $d$, and then the electrical resistance $R$ and temperature $T$ were calculated. A Keithley 2182A nanovoltmeter was used to have a direct view of the imbalance of the bridge, while this signal was recorded by means of a NI cDAQ-9178 data acquisition system with a NI 9239 voltage measurement unit at a 2 kHz acquisition rate. For each experiment, up to 100 identical approaches were performed and the resulting curves were averaged to obtain a better signal-to-noise ratio. A sliding average was also applied to remove the 50 Hz electrical noise.

In order to determine the emitter-cell thermal conductance during an approach, we measured the electrical resistance variation of the emitter and used the calibration data to calculate its temperature. The resistance measurements were first averaged before calculating the temperature and the conductance (SI and Ext. Data Fig. 1b). Variations of total thermal conductance $G_{tot} = \frac{P}{\theta}$, where $P$ is the electrical power fed to the probe and $\theta = T - T_{amb}$ the average temperature increase, are small. Thus the radiative thermal conductance in the near field $G_{NF}$, being exactly the variation $\Delta G_{tot}(d)$ which depends on the emitter-cell distance $d$, can be calculated from a logarithmic derivation: $\frac{\Delta G_{tot}}{G_{tot}} = \frac{\Delta P}{P} - \frac{\Delta \theta}{\theta} = \frac{\Delta R}{R} + \frac{2\Delta I}{I} - \frac{\Delta \theta}{\theta}$. With the reference temperature taken at the largest distance ($T_{ref}$), the temperature coefficient ($\alpha$), the emitter current and the reference current at the largest distance (respectively $I$ and $I_{ref}$), the following equation can be derived:

$$G_{NF} = G_{tot} \left[ (T_{ref} - T)(\frac{1}{T - T_{amb}} - \alpha) + 2\frac{I - I_{ref}}{I} \right] \quad (1)$$

where $G_{tot}$ is measured at the largest distance. Depending on the working temperature, $\alpha$ and ($I$-$I_{ref}$) are respectively positive/negative for $T < T_{Rmax}$, and negative/positive for $T > T_{Rmax}$ because the temperature drop during an approach can result in a decrease or an increase of electrical resistance. In Ext. Data Fig. 1b, the temperature drop close to contact is seen as a resistance increase due to the working temperature (≈ 732 K), which is higher than $T_{R\max}$. Finally, it must be noted that $G_{NF}$ is equal to 0 at the largest distances because it represents only the evanescent wave contribution to the radiative heat transfer, which is considered negligible at such distances.

Since radiative measurements can be noisy, they require averaging over many approach curves (see also SI and Ext. Data Fig. 1). At room temperature approximately 100 approach curves were considered for averaging. At low temperature (Fig. 3a), seven curves were used and consequently the signal-to-noise ratio became small for distances larger than 3 µm (shaded region). The value of the background level was estimated by averaging the measured flux over a distance range of 0.3 µm around the 4 µm position. This sets the main uncertainty on the exact flux determination.

**Design and fabrication of the photovoltaic cells**

The photovoltaic cells (Fig. 1c) were specifically designed and fabricated to demonstrate the near-field enhancement of the electrical power[15,42]. For the semiconductor material, we chose indium antimonide (InSb) for its low bandgap energy of 0.23 eV ($\lambda_g$ = 5.3 µm) at 77 K, thus allowing the conversion of low-energy infrared radiation (Fig. 1d).

We first analyzed numerically the doping levels and layer thicknesses of the InSb p-n junction to be able to obtain the largest electrical power generation from the cell. Second, we fabricated the optimum cells and measured their photovoltaic (PV) response in the far field using different emitter temperatures. Cells were made with different active area diameters of 20, 40, 80 and 160 µm in order to study the influence of the illuminated area on electrical power generation[15]. In order to investigate the impact of cell parameters on conversion performances, two p-doping levels of $10^{17}$ and $10^{18}$ cm$^{-3}$, and two substrate thicknesses of 200 and 500 µm, were selected. The cells were glued with silver paste on a gold-coated chip carrier and connection from the front and back contacts of the cells to the chip carrier was made using wire bonding.

In the present work, we added one step to the previously reported fabrication process[15]. Passivation is required to avoid accumulation of majority carriers at the side walls of the mesa and the associated current that lowers performances of the junction. The extra step consisted in passivating 20 µm of the circular mesa structure (10 µm from inside of circle and 10 µm from the etched area) using UV-insulated then cured AZ1518 resist. By means of a plasma-enhanced chemical vapor deposition (PECVD) method at 200 °C, all surfaces, except the cell active area, were then covered with a dielectric material such as $SiO_2$ or $Si_3N_4$, which allows preventing shortcuts between contacts and makes it possible to wire-bond. Top views of the resulting cells are shown in Ext. Data Fig. 7.

**Setup of the experiment**

The experimental setup (Fig. 1a) was assembled in a vacuum chamber where the pressure was decreased to $P \approx 10^{-6}$ mbar, using only an ion pump at that threshold to avoid vibrations. The chip carrier of the cell was glued with silver paste on the cold finger of a liquid-helium cryostat. The cell position was fixed in the experiment. The emitter was initially placed at approximately 2 mm above the surface of the cell and could be moved in $x$-, $y$- and $z$-axis directions by a custom piezoelectric positioning system having ranges of 3, 3 and 2.5 mm, respectively. Positioning of the emitter above the centre of the active area of the cell was done in three steps. First, the position of the emitter was controlled using piezoelectric positioners and a microscope camera (Dino-Lite Edge 3.0 AM73115MTF) with a long working distance of 15 cm for a 40x magnification located outside the vacuum chamber. The emitter was brought in contact with the chip at an arbitrary location, then retracted at a safe distance of around 100 µm and moved over the cell (see Ext. Data Fig. 8 for a top view). In a second step, the emitter was brought in contact with the cell and retracted at a distance $d$ smaller than 5 µm. There the fine positioning (step 3) was performed by measuring the short-circuit current ($I_{SC}$) of the cell according to the emitter displacement along the $x$ and $y$ axes, with the emitter heated at a temperature larger than 730 K. Once the maximum current is reached along one axis (see Ext. Data Fig. 8c), the maximum current position along the other axis corresponds to the centre of the active area. It is reminded that parallelization of the emitter and cell surfaces was not required due to the sphere-plane configuration, chosen on purpose in place of the plane-plane configuration.

**Detection of contact**

The mechanical contact between the emitter and the cell was detected simultaneously by two different methods. The temperature of the emitter was measured continuously according to the piezo displacement. At contact, there is a sudden drop of the temperature due to the contribution of heat conduction which becomes largely dominant over thermal radiation for the heat transfer between the emitter and the cell. The other method was based on the measurement of the cell current according to the displacement of the emitter. When the emitter is in contact, the performances of the cell harshly decrease due to the temperature increase[15] caused by the hot emitter. A significantly-lower photogenerated current is then measured. These two methods applied simultaneously were useful to check that the contact actually occurs in the active area of the cell. When some contact took place with the gold surface sides, the temperature drop of the emitter was still observed but the decrease of the photogenerated current was absent because the heated area does not contribute to the photovoltaic effect. More details on the analysis of the emitter-cell distance is provided in SI and Ext. Data Fig. 6.

**I-V characteristics measurement according to emitter-cell distance**

The current-voltage (I-V) characteristics of the photovoltaic cells were measured using a Keithley 2400 source measurement unit. A voltage scan was applied to the cell and the resulting current was measured (Fig. 2). For the dark configuration, we used a cooled radiative shield that was mounted around the sample to prevent the active area of the cells from receiving ambient temperature radiation. In order to get the electrical power at the maximum power point according to the emitter-cell distance, I-V curves must be measured as a function of distance. For stability purposes, it was chosen to perform continuous approaches of the emitter while measuring the current produced by the cell for a fixed voltage, and then to repeat this procedure for different voltages. After performing approaches at voltages ranging from 0 to around 140 mV, we obtained a series of current-distance curves at different cell voltages that could be used to plot the current-voltage curves at different distances. Then we calculated the photovoltaic power at each distance and determined the maximum value so as to plot the maximum electrical power according to distance. We also used another method based on the superposition principle, where the I-V curve of the cell is considered to keep the same shape but is shifted in short-circuit current depending on the illumination level. With this method, we measured a reference I-V curve under ambient illumination, then we measured only the short-circuit current $I_{SC}$ (current at $V_{cell} = 0$) according to distance. The reference I-V curve was then shifted by plotting the (I-$I_{SC}$)-V curves at each distance. Then the maximum electrical power could be calculated. This method was much faster to perform experimentally because it required approach curves at only one cell voltage, but in this case the maximum power was deduced from the shifted reference curves and not directly measured. Validation of the superposition principle is provided in Ext. Data Fig. 9.

**Near-field conversion efficiency measurement**

The setup allows the simultaneous measurement of the electrical power generated by the cell and the radiative power exchanged between the emitter and the cell in the near field. In order to identify only the near-field contribution to electrical power generation, the theoretical evolution of power according to distance was calculated using the analytical expression of the sphere-disc view factor. We used the electrical power measured at the largest distance in an approach curve similar to Fig. 3b as the fitting parameter, assuming that the generated power was only due to the far-field contribution. The difference between the power calculated with the view factor and the total measured power corresponds to the near-field electrical power. Note that our definition of the far-field contribution is that related to the view-factor theory, and that it can depart from an exact calculation of the contribution of the propagative waves in the selected geometry by means of fluctuational electrodynamics. In case of a slight error on the identification of the far-field contribution from the experimental data (Fig. 3), our estimation is a lower bound to the near-field contribution. For the different distances, the ratio between the near-field electrical power and the near-field exchanged radiative power gave the conversion efficiency for the near-field contribution only.

**Methods reference**

54. Spieser, M., Rawlings, C., Lörtscher, E., Duerig, U. & Knoll, A. W. Comprehensive modeling of Joule heated cantilever probes. *J. Appl. Phys.* **121**, (2017).

**Acknowledgments**

Financial support by the French National Research Agency (ANR) under grant No. ANR-16-CE05-0013 and partial funding by the French "Investment for the Future" program (EquipEx EXTRA ANR-11-EQPX-0016 and IDEXLYON ANR-16-IDEX-0005) and by the Occitanie region are acknowledged. C.L. and P.O.C. thank C. Ducat, N. Pouchot and A. Buthod for help in the design of the setup, P. Mangel and D. Renahy for experimental support, and S. Gomes. P.O.C thanks S. Volz for equipment handover.

**Author contributions**

R.V., P-O.C, J-P.P., E.T. and T.T. conceived and supervised the work. J-P.P did the MBE growth of the InSb material and D.C. fabricated the PV cells. C.L. fabricated the emitter and performed the experiments. C.L. and R.V. performed the simulations. The manuscript was written by C.L., P-O.C and R.V. with comments and inputs from all authors.

The authors declare no financial competing interest.

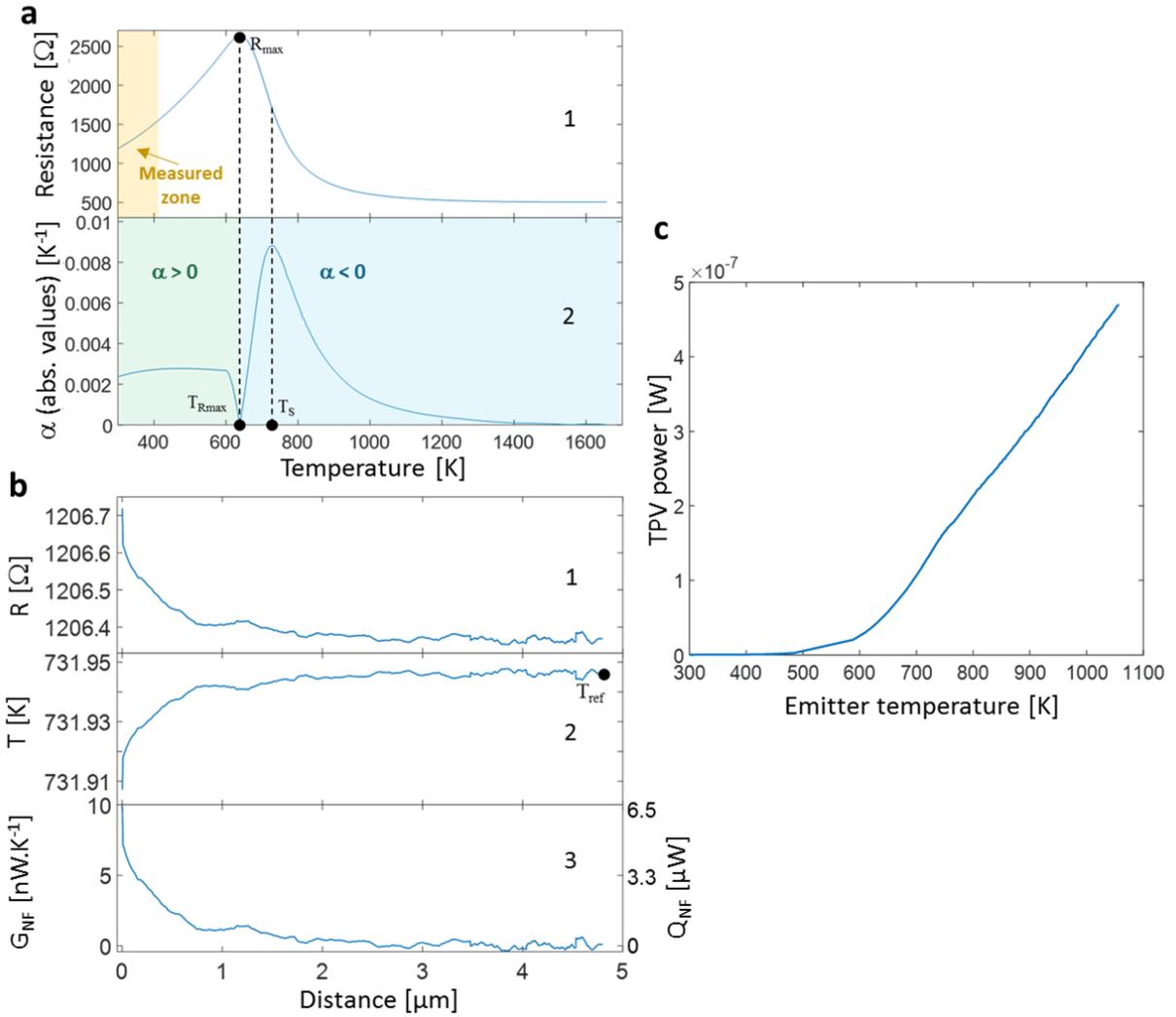

**Extended Data Figure 1 | Determination of the near-field radiative thermal conductance and impact of emitter temperature.** (a) Calibration of the emitter temperature: (Step 1) Electrical resistance and (Step 2) temperature coefficient of the electrical resistivity ($\alpha = \frac{1}{R}\frac{dR}{dT}$) of the emitter according to the oven temperature. (b) Application of the calibration for determining the thermal conductance**:** (Step 1) Electrical resistance, (Step 2) temperature and (Step 3) near-field radiative thermal conductance as a function of distance to the contact. The emitter-sample power exchanged is also indicated. The electrical resistance measurements were averaged over 16 curves. (c) Impact of the emitter temperature on the near-field thermophotovoltaic (TPV) power output, measured for a photovoltaic cell with diameter 20-μm at $d \approx 5$ μm.

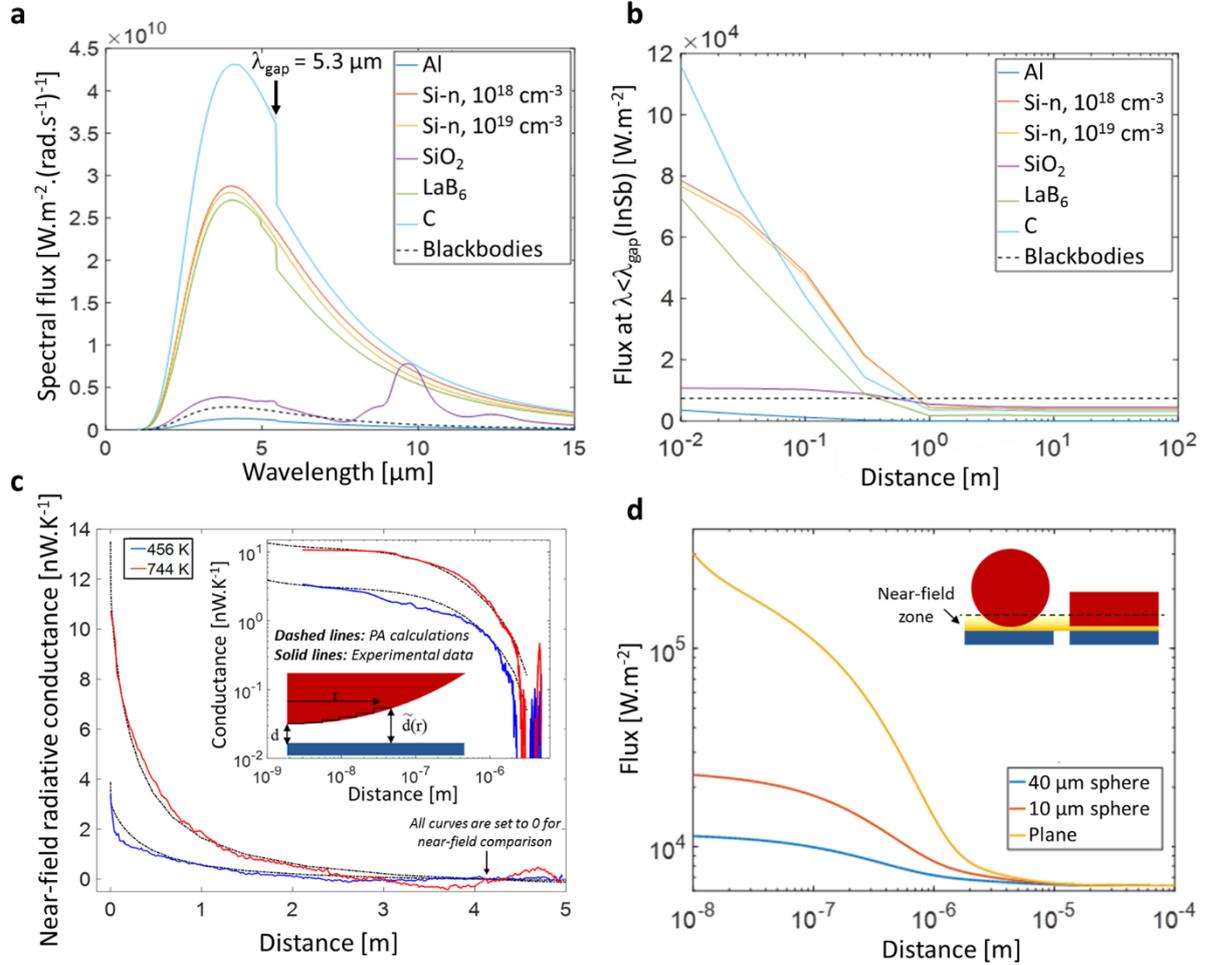

**Extended Data Figure 2 | Near-field radiative power exchanged for different emitters in the vicinity of semi-infinite planar bodies.** (a) Spectral flux between two semi-infinite planar media. Calculations for different emitters at 732 K and an InSb sample at 77 K with an emitter-sample distance $d$ = 10 nm. (b) Integrated flux for radiation wavelengths useful for photocurrent generation ($\lambda < \lambda_{gap}$), as a function of distance. See SI Sec. 1. (c) Near-field thermal conductance as a function of distance, for two spherical graphite emitter temperatures and an InSb sample at room temperature. Inset: logarithmic plot. Plain lines are experimental data and dashed lines are predictions from the Proximity Approximation. The schematic reminds the principle of the Proximity Approximation (use of the infinite parallel-surfaces heat transfer coefficient and integration as a function of height, see SI Sec. 2). (d) Power divided by the projection over the surface area for two different geometries (planar or spherical emitter). The graphite emitter temperature is 732 K and the InSb sample is at 77 K. The fluxes for the sphere-planar surface geometry are computed by means of the Proximity Approximation.

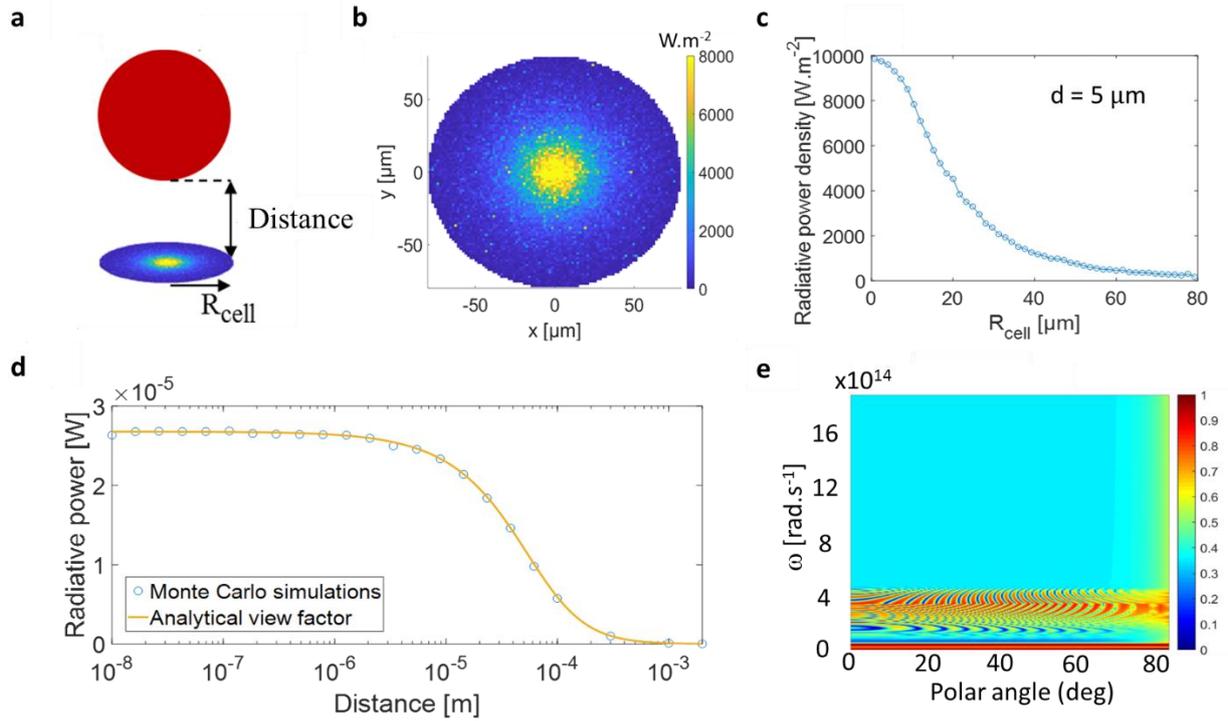

**Extended Data Figure 3 | Calculations of the contribution of the propagative waves to the radiative power according to distance** (see SI Sec. 4). (a) Schematic of the Monte Carlo computation of the propagative photon contribution to the transfer between the hot sphere and the cold receiver (see SI). (b) Local radiative power density deposited by a 40 µm graphite sphere at 732 K on a flat InSb semi-infinite cylinder of diameter 160 µm at 77 K, for an emitter-cell distance equal to 5 µm. The cylinder represents the cell. (c) Radiative power density as a function of cell radius. (d) Evolution of the radiative power calculated using the view factor as a function of distance, compared with the Monte Carlo simulations. (e) Spectral and directional specular reflectance of an InSb cell extended infinitely in the lateral directions and consisting of four layers (p-doped, $N_a = 10^{17}$ cm$^{-3}$, $t_p = 0.5$ µm / n-doped, $N_d = 10^{15}$ cm$^{-3}$, $t_n = 2.5$ µm / n-doped substrate, $N_{d,sub} = 4\ 10^{17}$ cm$^{-3}$, $t_{sub} = 500$ µm / gold, $t_{brl} = 200$ nm). These data, computed by electromagnetic means, are fed in the Monte Carlo computation of the far-field contribution.

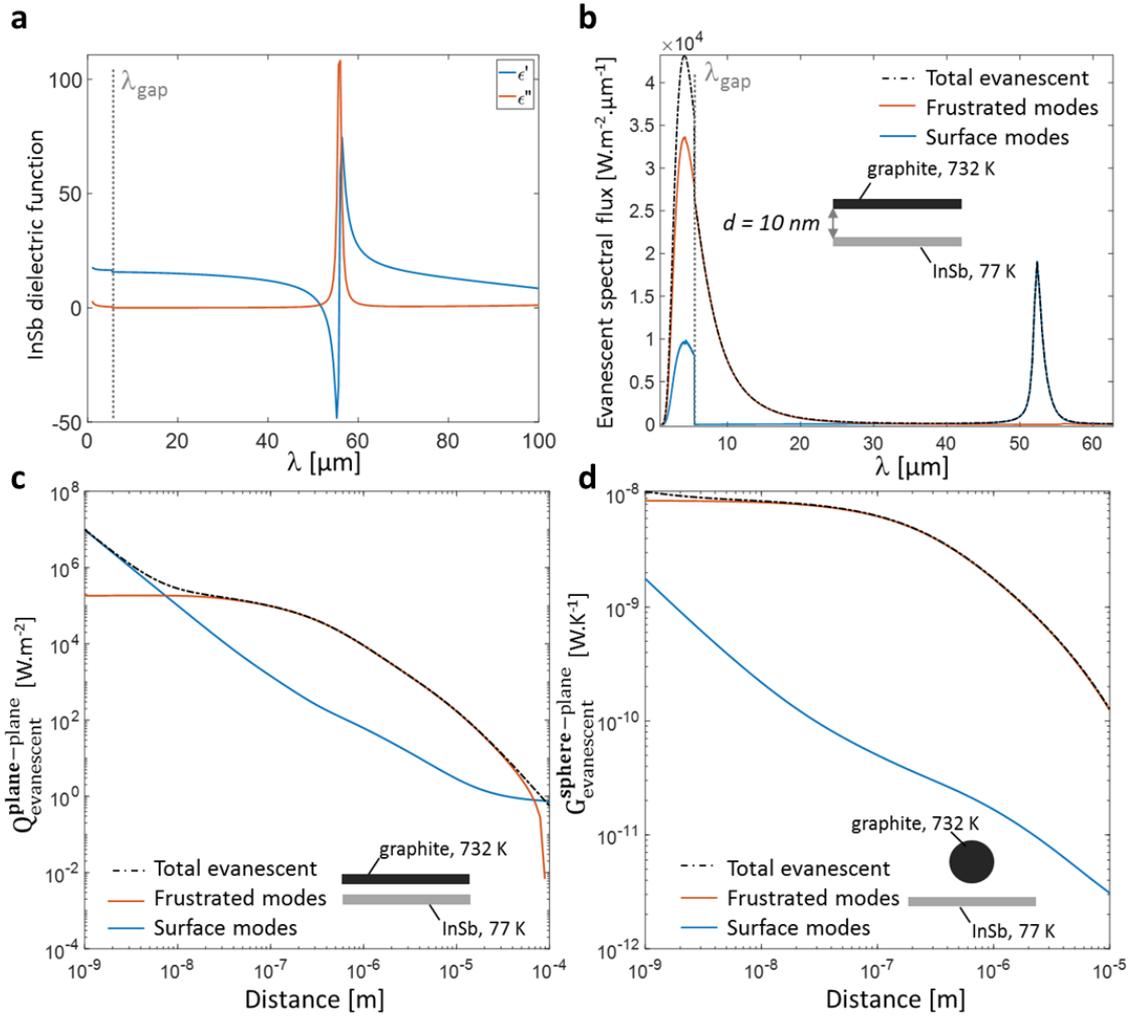

**Extended Data Figure 4 | Contribution of the frustrated and surface modes to the evanescent radiative heat transfer**. The emitter temperature is 732 K and the InSb semi-infinite planar medium receiver is at 77 K (a) Real (blue) and imaginary (red) parts of the dielectric function of InSb at 77 K (p doping level of $10^{18}$ cm$^{-3}$). (b) Evanescent component of the spectral flux between two semi-infinite planar media separated by a 10 nm vacuum gap. (c) Evanescent flux as a function of the distance between two semi-infinite planar media. (d) Evanescent conductance as a function of distance, for a 37.5 μm graphite sphere and an InSb semi-infinite planar medium. See also SI Sec. 5.

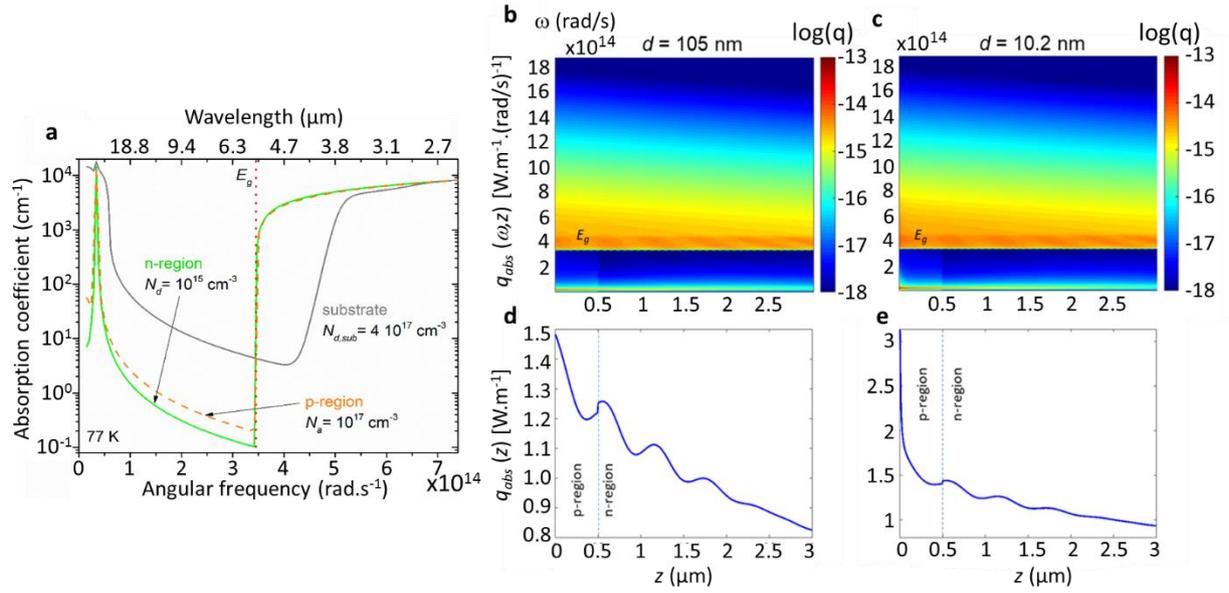

**Extended Data Figure 5 | Radiative power absorbed as a function of frequency and depth, for sphere-cell distances of 105 nm (b,d) and 10.2 nm (c,e) and a substrate thickness of 500 μm**. (a) Spectral absorption coefficient of indium antimonide for different doping levels considered for these calculations. (b,c) Local spectral radiation power absorbed as a function of depth ($z$) inside the cell. (d,e) Local radiation power absorbed as a function of depth. The results are computed with the Proximity Approximation for both evanescent and propagative contributions, and with the following parameters: graphite sphere of diameter 40 μm at 773 K, InSb cell at 77 K made of four layers (p-doped, $N_a = 10^{17}$ cm$^{-3}$, $t_p = 0.5$ μm / n-doped, $N_d = 10^{15}$ cm$^{-3}$, $t_n = 2.5$ μm / n-doped substrate, $N_{d,sub} = 4\ 10^{17}$ cm$^{-3}$, $t_{sub} = 500$ μm / gold, $t_{brl} = 200$ nm). Surface modes decay quickly from the surface at $d = 10$ nm, and are not present at 100 nm. See also SI Sec. 5.

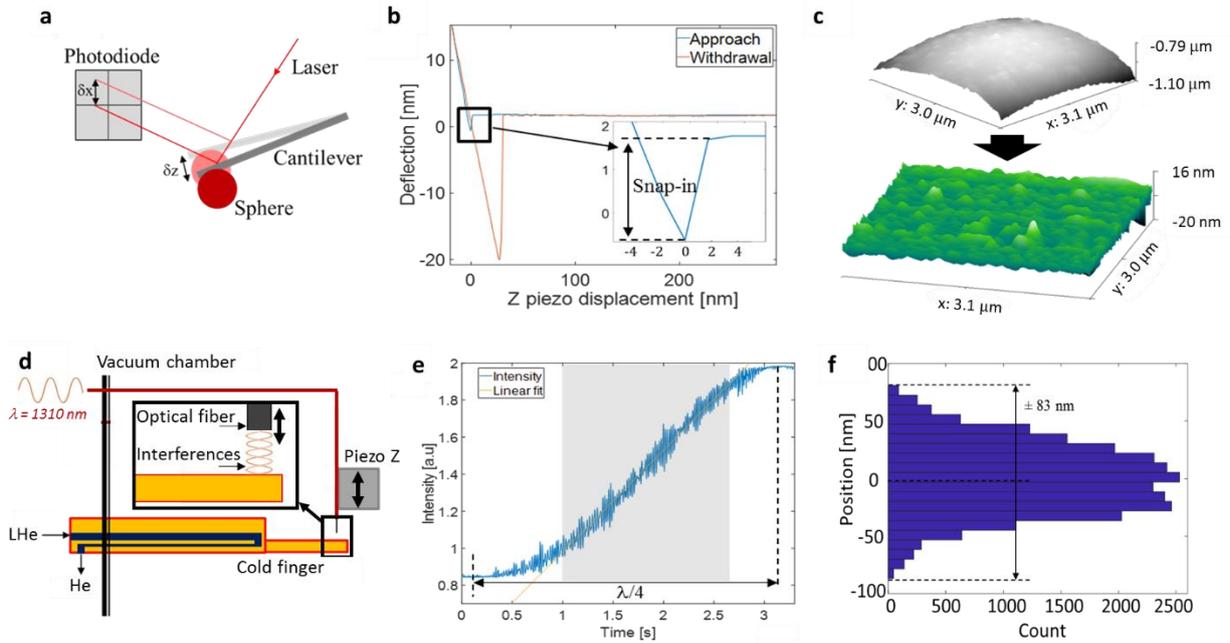

**Extended Data Figure 6 | Parameters influencing the determination of the emitter-cell distance at contact,** *i.e.* **the snap-in distance associated to the jump at contact (a,b), the sphere material roughness (c), and sample vibrations (d-f)** (see also SI Sec. 6). (a) Schematic of the cantilever deflection measurement setup using a quadrant photodiode system. (b) Cantilever deflection as a function of the displacement during an approach (blue) and a withdrawal motion (red). The emitter is out of contact when the curve is flat and is in contact when the slope is steep. The data indicate an uncertainty of few nanometers at maximum. (c) Topographic image of the sphere obtained by scanning the emitter with an atomic force microscopy tip, before and after spherical shape subtraction. This allows determining the roughness of the emitter involving an rms roughness of 5 nm and peaks of maximal height around 30 nm. (d) Schematic of the interferometric vibration measurement setup allowing to measure the variation of position of the cell with respect to the cantilever basis. (e) Interferometric signal allowing to calculate the vibration amplitudes. The linear domain is represented by the shaded area. (f) Histogram of the cold finger positions, indicating a maximal vibration amplitude of about 80 nm.

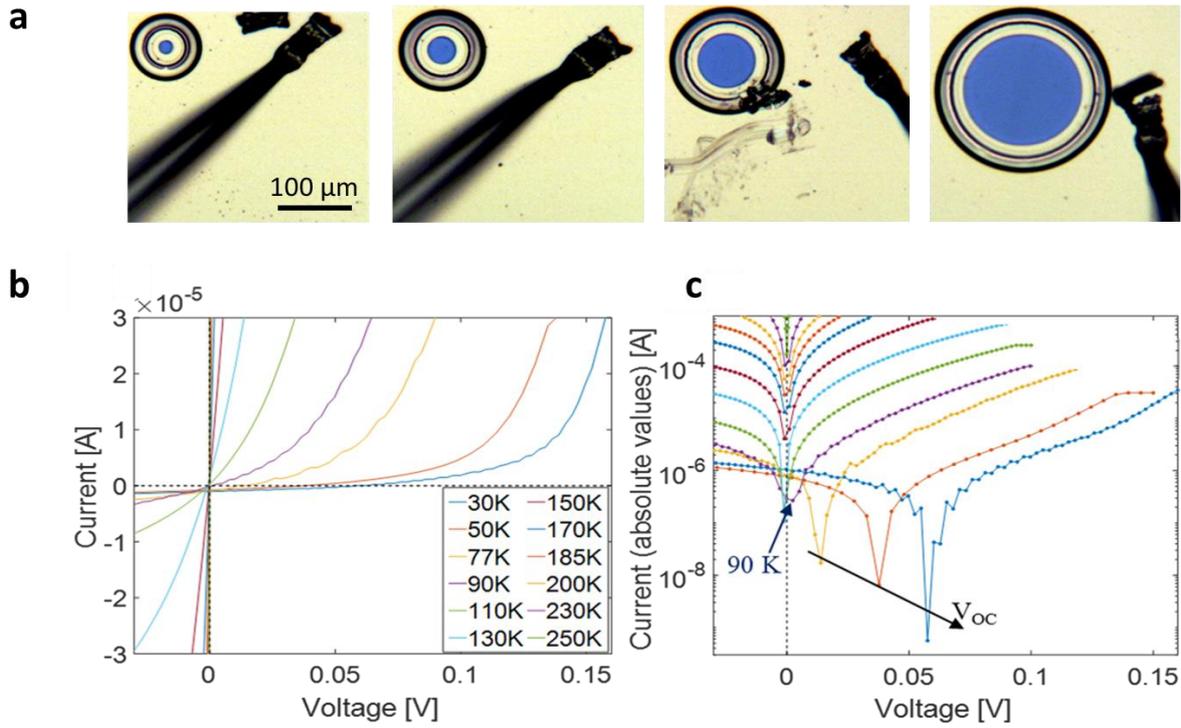

**Extended Data Figure 7 | Images and characteristics of the cells.** (a) Top view of the InSb photovoltaic cells with different active area diameters. (b) I-V curves at different temperatures. (c) I-V curves at different temperatures with the current represented in absolute value and using a logarithmic scale. See SI Sec. 7.

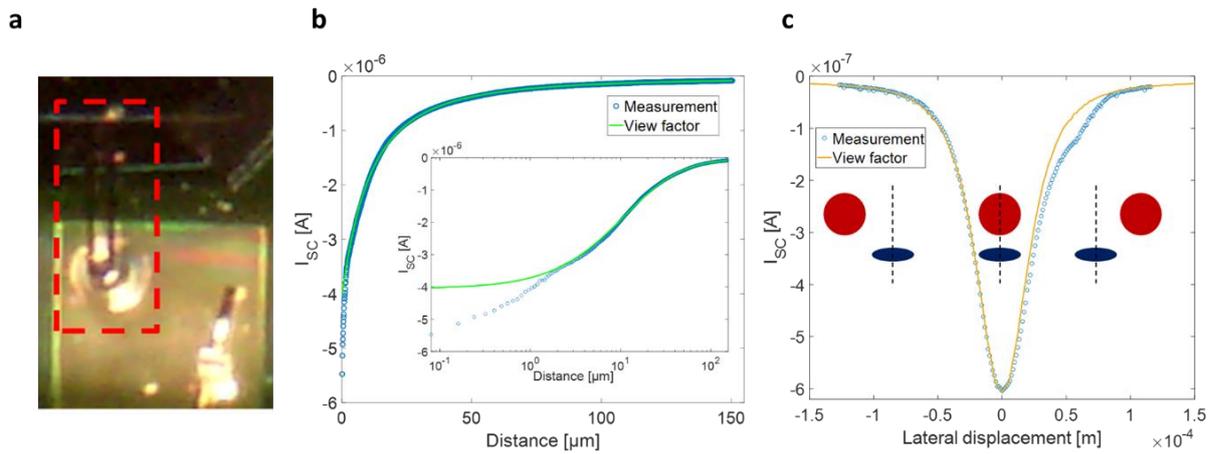

**Extended Data Figure 8 | Variation of the short-circuit current as a function of the position of the emitter.** (a) Optical microscopy image of the setup involving the chip where a thermophotovoltaic cell (disc where there is no gold) is seen. Red box: View from top of the scanning thermal microscopy probe on which the spherical emitter is glued, positioned over a cell. (b) Short-circuit current as a function of the vertical displacement of the emitter and comparison with the view factor. Same data as that of the inset of Fig. 3b, but in linear scale. (c) Short-circuit current as a function of the lateral displacement of the emitter and comparison with the view factor. See SI Sec. 8.

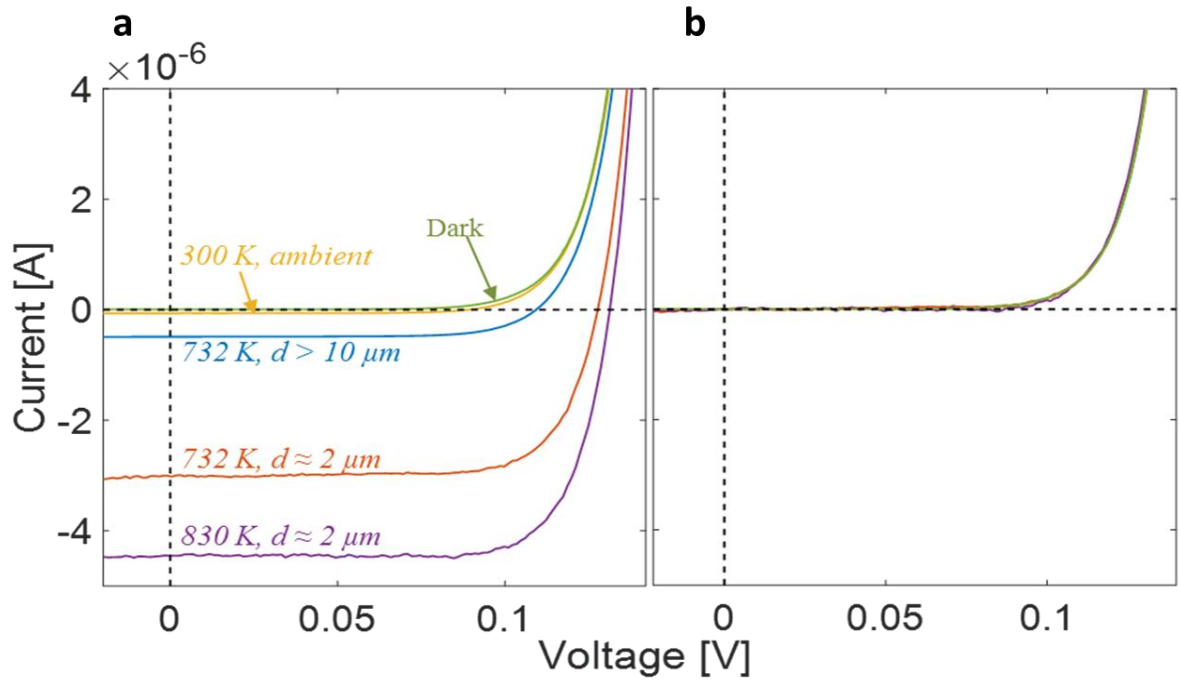

**Extended Data Figure 9 | I-V curves of a cell having a 20 μm active area diameter in the dark and under different illumination levels.** (a) I-V curves under illumination from different distances and temperatures. (b) Curves shifted in current: the short-circuit current of each curve is set equal to zero. All curves superimpose, as expected from the superposition principle (see SI Sec. 9).

| Method | $T_{emitter}$ [K] | p doping level [$cm^{-3}$] | Substrate thickness [μm] | ΔT [K] | $D_{active\ area}$ [μm] | Max TPV power [μW] | Max TPV power density [kW.m$^{-2}$] | Max NF-TPV power [μW] | Max NF-TPV power density [kW.m$^{-2}$] | Max NF radiative flux [μW] | Estimated FF radiative flux [μW] | Estimated FF radiative power density [kW.m$^{-2}$] | Estimated total efficiency [%] (uncertainty see text) | Determined NF efficiency [%] (uncertainty see text) | NF-TPV enhancement factor |
|---|---|---|---|---|---|---|---|---|---|---|---|---|---|---|---|
| ≠ V | 732 | $10^{17}$ | 500 | 655 | 80 | 2.96 | 0.60 | 0.40 | 0.08 | - | 16.8 | 3.34 | | | 1.2 |
| ≠ V | 732 | $10^{17}$ | 500 | 655 | 20 | 0.55 | 1.80 | 0.31 | 0.99 | - | 3.39 | 10.8 | | - | 2.3 |
| Sup. pr. | 732 | $10^{17}$ | 500 | 655 | 20 | 0.55 | 1.80 | 0.32 | 1.02 | 5.3 | 3.39 | 10.8 | 6.3 | 6.0 | 2.4 |
| Sup. pr. | 830 | $10^{17}$ | 500 | 655 | 20 | 1.08 | 3.40 | 0.70 | 2.24 | 8.8 | 6.12 | 19.5 | 7.2 | 7.9 | 2.9 |
| Sup. pr. | 732 | $10^{17}$ | 500 | 753 | 160 | 4.82 | 0.24 | 1.11 | 0.06 | - | 22.2 | 1.10 | | - | 1.3 |
| ≠ V | 732 | $10^{17}$ | 500 | 655 | 160 | 5.45 | 0.27 | 0.99 | 0.05 | 7.6 | 22.2 | 1.10 | 18.3 | 13.0 | 1.22 |
| Sup. pr. | 732 | $10^{17}$ | **200** | 655 | 20 | 1.28 | 4.10 | 1.05 | 3.20 | 6.5 | 3.20 | 10.2 | 12.2 | 16.2 | 4.8 |
| Sup. pr. | 900 | $10^{17}$ | **200** | 823 | 20 | 1.90 | 6.00 | 1.41 | 4.50 | 14.1 | 7.84 | 25.0 | 8.7 | 10.0 | 4.0 |
| ≠ V | 732 | $10^{17}$ | **200** | 655 | 20 | 1.25 | 3.97 | 0.97 | 3.10 | 6.9 | 3.20 | 10.2 | 12.4 | 14.1 | 4.6 |
| Sup. pr. | 732 | $10^{18}$ | 500 | 655 | 20 | 2.36 | 7.50 | 1.96 | 6.25 | - | 3.44 | 10.9 | | - | 5.9 |
| ≠ V | 732 | $10^{18}$ | 500 | 655 | 20 | 2.37 | 7.50 | 1.79 | 5.70 | | 3.44 | 10.9 | | | 4.1 |

**Extended Data Table 1 | Results of the near-field thermophotovoltaic experiments for different studied configurations** (see also SI Sec. 10).

# SUPPLEMENTARY INFORMATION

**Outline**

**1. Supplementary Methods**

**2. Supplementary Discussion**

# 1. Supplementary Methods

1. <u>Selection of the emitter material</u>

In order to choose an efficient emitter material that matches the optical properties of InSb, we performed radiative heat transfer calculations between two semi-infinite planar media labelled 1 and 3 separated by a vacuum gap labeled 2. The spectral heat flux emitted by one of these media at temperature $T$ and absorbed by the second one is the sum of the propagative and evanescent contributions $q = q_{prop} + q_{evan}$ and can be calculated based on the electric properties, here the complex dielectric function $\varepsilon$ which depends on the angular frequency $\omega$, of the emitter and the receiver[9]:

$$q_{evan}(\omega) = \frac{1}{\pi^2} \frac{\hbar\omega}{e^{\frac{\hbar\omega}{k_B T}}-1} \int_{k_\parallel=k_0}^{\infty} k_\parallel e^{-2Im(k_{2\perp})d} \sum_{i=TE,TM} \frac{Im(r_{21}^i)Im(r_{23}^i)}{|1-r_{21}^i r_{23}^i e^{2ik_{2\perp}d}|^2}, \quad \text{(S1a)}$$

$$q_{prop}(\omega) = \frac{1}{4\pi^2} \frac{\hbar\omega}{e^{\frac{\hbar\omega}{k_B T}}-1} \int_{0}^{k_\parallel<k_0} k_\parallel \sum_{i=TE,TM} \frac{(1-|r_{21}^i|^2)(1-|r_{23}^i|^2)}{|1-r_{21}^i r_{23}^i e^{2ik_{2\perp}d}|^2}, \quad \text{(S1b)}$$

We remind that $\hbar\omega$ is the energy of a photon, $k_B$ is Boltzmann's constant, and the Fresnel coefficients are defined as

$$r_{21}^{TE} = \frac{k_{2\perp}-k_{1\perp}}{k_{2\perp}+k_{1\perp}}, \quad \text{(S2a)}$$

$$r_{21}^{TM} = \frac{k_{2\perp}\varepsilon_1 - k_{1\perp}\varepsilon_2}{k_{2\perp}\varepsilon_1 + k_{1\perp}\varepsilon_1}, \quad \text{(S2b)}$$

where $k_\parallel$ is the part of the wavevector $k_0 = \omega/c$ which is parallel to a flat interface and $k_{i\perp} = \sqrt{\varepsilon_i k_0^2 - k_\parallel^2}$. $c$ is the velocity of light. The total net flux is the difference between that emitted by the hot body and absorbed by the cold one, and that emitted by the cold body and absorbed by the hot one.

Different emitter materials were considered at 732 K and coupled with InSb at 77 K. The resulting spectral fluxes are presented in Ext. Data Fig. 2a. We observed that in near-field conditions at $d =$ 10 nm, graphite is an excellent emitter because the spectral flux exchanged with InSb is enhanced by more than one order of magnitude compared to the blackbody limit at wavelengths useful for photocurrent generation ($\lambda < \lambda_{gap}$). The integrated flux over such wavelengths according to distance is also studied (Ext. Data Fig. 2b). At $d > 100$ nm, doped silicon is slightly better than graphite but tends to level off at the lowest distances. As a result, graphite was chosen with the aim of studying the sub-100 nm regime for thermophotovoltaic conversion. Ideally, a better emitter material would be supporting surface phonon polaritons at wavelengths close to $\lambda_{gap}$(InSb) at 77 K, in order to enhance by several orders of magnitude the radiative heat transfer and the electrical power generation in the near field.

2. Sphere-plane vs. plane-plane configurations

We chose a sphere-plane instead of a plane-plane configuration to avoid any parallelization issue. A drawback is that the near-field enhancement is strongly reduced since a lower fraction of the emitter area is in the near-field zone compared to a planar one at the same distance (Ext. Data Fig. 2d). In order to estimate the radiative flux between a spherical emitter and a semi-infinite planar medium, we used the proximity approximation (PA, also called Derjaguin approximation)[55] for computing the contribution of the near-field flux. A flux database was calculated for the case of the plane-plane geometry from 1 nm to 100 µm for the selected temperature. The principle of the PA is that the spherical shape is approximated as the sum of small planar elements and the sphere-plane flux or conductance can be calculated by integrating the local plane-plane flux weighted by the local perimeter over the half sphere (see schematic in the inset of Ext. Data Fig. 2c):

$$G_{sphere-plane}(d,T) = \int_0^{R_{sphere}} G_{plane-plane}[\tilde{d}(r),T] 2\pi r dr, \qquad (S3)$$

where $\tilde{d}(r) = d + R_{sphere} - \sqrt{R_{sphere}^2 - r^2}$. We calculated the radiative heat flux between a graphite sphere with diameter 40 µm at 732 K and an InSb plane at 77 K and compared it with a planar emitter having the same surface area (Ext. Data Fig. 2d). As expected, the enhancement due to the contribution of the evanescent waves at low distances is much weaker with the spherical emitter (more than one order of magnitude). With a smaller sphere, a larger fraction of the area is in the near-field zone, so the exchanged power value varies more strongly with distance. However, the smaller the sphere, the smaller the exchanged power to be measured. Thus we chose a diameter of 40 µm as a good compromise between near-field enhancement and power detectability. Consequently, with the sphere-plane geometry, a weaker enhancement of the TPV power in the near field is expected in comparison with the plane-plane geometry.

3. Comparison between near-field thermal radiative measurements and the proximity approximation

Before making measurements with the cells at low temperature, we validated our experimental setup by measuring the near-field thermal radiation exchange at room temperature between different emitters and substrates. Ext. Data Fig. 2c reports on measurements between a graphite emitter and a flat InSb substrate. In order to obtain a better signal-to-noise ratio, about 100 approach curves were accumulated and averaged in the present cases. The contact point shifts between the approaches due to some long drift and is therefore tracked before averaging the data by searching for the distance where the electrical resistance varies abruptly. The effect of the emitter temperature can be observed: the near-field radiative conductance increases by half a decade when the emitter temperature is varied from 456 K to 744 K.

A comparison with the prediction of the Proximity Approximation (Eq. (S3)) is also shown in Ext. Data Fig. 2c. It is observed that there are some little differences, ascribed to the fact that the Proximity Approximation is more accurate at small distances than at micrometer-scale ones. At very small distances (in the sub-200 nm regime), where the position is determined less accurately (see Sec. 6), a lower slope is observed in the logarithmic plot for both the experimental data and the predictions.

4. Contribution of the propagating modes with distance

The far-field radiative flux was estimated based on two different approaches. First, the macroscopic theory of view factors was used by assuming diffuse isotropic surfaces. The analytic expression of the sphere-disc view factor depending on the cell and sphere radii $R_{disc}$ and $R_{sphere}$, and the sphere-cell distance $d$, reads as follows[56]:

$$F_{s \to d} = \frac{1}{2} \left( 1 - \frac{1}{\sqrt{1 + \left(\frac{R_{disc}}{d + R_{sphere}}\right)^2}} \right) \tag{S4}$$

Due to the flatness of the surfaces, it could be argued that the surfaces are specular and not diffuse. A second method was then implemented, using a Monte Carlo ray tracing numerical approach (see principle on Ext. Data Fig. 3a). For precise calculations, we took into account the properties of graphite and the materials constituting the InSb cell, such as emissivity and reflectance, as a function of emission angle. In this method, rays are launched from the emitter into all directions. A fraction of the total power emitted by the sphere is attributed to each ray depending on the spectral distribution of the radiative flux at the emitter temperature. Then calculations are made to check if the rays coming from the sphere are crossing the surface of the cell. If so, the trajectory is extended by considering the angle of incidence, and the power lost by the ray (absorbed by the cell) is calculated using the reflectivity of InSb. Then we check if the extended trajectory crosses the surface of the sphere. Using the same principle as before, the power loss and the reflected trajectory are calculated. In this method, we considered up to 3 reflections on the sphere for the same ray, because it was estimated that the number of remaining rays and their energy after such a number of reflections were negligible. The results of the Monte Carlo method which are shown are for a spherical emitter with diameter of 40 µm and a finite disc with different diameters as the receiver. The total flux absorbed by the disc, which is representative of the cell, is provided by a map of the power density (Ext. Data Fig. 3b) and a power density profile integrated as a function of cell radius (Ext. Data Fig. 3c). We observe in the figures that the flux decays for radii larger than 20 µm, so it is not necessary to use cells with a too large radius as most of the surface would not be much illuminated by the sphere. We could also calculate the evolution of the radiative power deposited on the cell as a function of distance (Ext. Data Fig. 3d). We can see that for distances smaller than ≈ 3 µm, variations of the flux are very small so most of the measured enhanced power will be coming from the contribution of the evanescent waves in the near field. We compared the Monte Carlo simulations of the specular bodies considering $10^6$ rays with the analytical expression of the diffuse-emission view factor and found a very good agreement between the results. The specularity of the surface is therefore not key to the computation of the exchanged power. Taking into consideration this result and because the Monte Carlo simulations need a significant computational time, the propagating mode contribution was calculated for a single distance using the Monte-Carlo method. The flux at other distances was deduced from both this value and the analytical expression of the view factor. Note that the reflectance of the cell, which is a multilayered structure, were first calculated by electromagnetic means and then included in the Monte Carlo approach for the estimation of the propagative-wave contribution (see Ext. Data Fig. 3e).

5. Numerical analysis of the near-field modes and radiative absorption as a function of depth

It is key to study the contribution of the different evanescent modes to the heat flux, *i.e.* frustrated and surface modes. Frustrated modes undergo total internal reflection at an interface, they are propagative in the medium and evanescent in vacuum. Surface modes are evanescent both in vacuum and in the medium. The dielectric function of InSb (Ext. Data Fig. 4a) allows for resonant surface waves (phonon-polaritons) around the wavelength λ = 55 µm, which is by far larger than the wavelength corresponding to the bandgap of InSb at 77 K (5.3 µm) and $\lambda_{Wien}$ (3.96 µm at 732 K).

We consider a semi-infinite medium of InSb as the receiver. We first analyze the plane-plane configuration. In Ext. Data Fig. 4b, computed for $d$ = 10 nm (a very small distance with respect to possible applications), we observe a peak in the spectral flux around 55 µm due to the phonon-polariton of InSb. The distance dependence shows that the evanescent component of the radiative heat transfer is dominated by the frustrated modes for approximately $d$ > 40 nm (Ext. Data Fig. 4c). For the sphere-plane configuration (Ext. Data Fig. 4d), this translates into a domination of the frustrated modes for distances $d$ > 10 nm. Knowing the uncertainty on distance close to the contact (see Sec. 6), this study indicates that, in our case, the near-field radiative heat transfer is fully dominated by the frustrated modes. The surface modes do not contribute to the electrical power generation by the cell in our experiment.

Let us now move to actual geometries and materials. The dependence of the dielectric function on doping concentration is reminded in Ext. Data Fig. 5a. The absorption coefficient of InSb is calculated[4] at 77 K for the p-region ($N_a = 10^{17}$ cm$^{-3}$) in the standard configuration used in the core paper, the n-region ($N_d = 10^{15}$ cm$^{-3}$), and the n-doped substrate ($N_{d,sub} = 4\ 10^{17}$ cm$^{-3}$). Some differences can be seen below the bandgap, where the doping has a strong influence. The penetration depth just above the bandgap is large, and the substrate is even more transparent. The spectral heat flux absorbed as a function of position in the cell is reported in Ext. Data Fig. 5b,c. It is computed with the Proximity Approximation (see Secs. 2-3) and the methods described in Ref. 57. Note that in contrast to other sections, the PA is also applied here for the propagative contribution, which is a stronger approximation but allows obtaining an estimation of the absorption as a function of depth. At a distance of ~100 nm (Ext. Data Fig. 5b), which is close to the minimum one achievable in our experiment (see Sec. 6), all the power absorbed in the junction (first 3 µm from the top of the cell, see Fig. 1 for the whole architecture) is located above the bandgap and there is no contribution of the polariton as suggested by Ext. Data Fig. 4d. Some interference patterns are observed in the junction. The power absorbed decays slowly as a function of depth, which calls for the introduction of a mirror as close as possible from the junction to allow for a second passage of the photons that have not been absorbed. We remind that the electron-hole pair generation rates is a direct translation of the absorbed power in the junction above the bandgap. Let us now analyze smaller distances ($d$ ~ 10 nm), which may be more difficult to implement in practical applications. Now the contribution of surface modes, which decay quickly from the surface, can be observed (Ext. Data Fig. 4c,d). Exploiting these modes requires positioning the np contact region close to the top surface and induces addition constrains. While this distance regime was not tested experimentally, our design is compatible with it.

## 6. Estimation of the sphere-cell distance close to contact

### a) Effect of the attraction forces

The distance when the sphere is few nanometers close to contact can also be modified by the attraction forces between the sphere and the surface. These forces are expected to bend the SThM probe cantilever and thus bring the sphere into contact. In order to quantify the distance where the snap-in occurs, we performed cantilever deflection measurements as a function of distance at room temperature for different emitter temperatures. The deflection was measured with a photodiode system in an NTMDT atomic force microscope (AFM) equipped with a moderate-vacuum chamber ($10^{-1}$ mbar). In this experiment, a laser was focused on the edge of a SThM probe with a sphere glued on the tip, and the reflection of the laser was observed with a quadrant photodiode. The cantilever deflection $\delta z$ could be measured by looking at the $\delta x$ displacement of the reflected laser on the photodiode (Ext. Data Fig. 6a). In out-of-contact position, the deflection of the probe is constant because there is no interaction between the probe and the sample. In the approach curve close to contact (Ext. Data Fig. 6b), the cantilever bends slightly due to the attraction forces. We estimated this bending to be around 2-3 nm. Thus it corresponds to the distance range not achievable with this experimental configuration. It can also be noticed that the adhesion forces, taking place when the probe is withdrawn, lead to a much larger deflection, greater than 20 nm.

### b) Effect of roughness

At low distances, the effect of surface roughness can become important. We performed roughness measurements both on the sphere and the cell using atomic force microscopy to acquire topographic images. By using the same principle as before, the deflection of an AFM probe was measured when the surface of the sample was scanned is $xy$ directions. The root mean square roughness $R_{RMS}$ and more importantly the maximum peak height and valley depth were determined from these measurements. For the sphere, an additional data processing was necessary as the surface is not flat: a spherical shape with the corresponding radius of curvature was subtracted from the measured data to obtain a flat topographic image (Ext. Data Fig. 6c). We found that the cells were almost perfectly flat with $R_{RMS} = 0.2$ nm whereas the graphite spheres had $R_{RMS} = 5.2$ nm with +29.9 nm and -28.7 nm as peak height and valley depth, respectively. These values mean that when we considered the sphere to be in contact with the cell for $d = 0$, the effective distance is most probably $d \approx 30$ nm between the surface of the cell and the mean spherical shape. This indicates that the study is restricted to distances larger than 30 nm.

### c) Vibrations of the cryostat

In order to cool the cells down to 77 K, we used a liquid helium (LHe) cryostat with a cold finger located inside a vacuum chamber (LHe was used for practical reasons but liquid nitrogen can also be used). The cold finger is 15 cm long inside the vacuum chamber so the continuous flow of LHe in the finger induces mechanical oscillations. In our case it is very important to know the amplitude of these vibrations because it corresponds to the minimum achievable emitter-cell distance. We used an Attocube LDM1300 interferometric module based on an IR laser with a 1310 nm wavelength fed through an optical fiber in order to measure the amplitude of the vibrations. The fiber was attached to the setup on the $z$-piezo where the emitter is usually located during the experiments, and was placed close to the surface of the cold finger (see Ext. Data Fig. 6d). Then we moved the fiber at a constant speed over a 5 µm range by doing a series of approach/withdrawal motions while the periodic interferometric signal was acquired. When the fiber moves by a distance λ corresponding to the wavelength of the laser, the optical path is modified by 2λ so the period of the interferometric signal corresponds to a λ/2 displacement. We can deduce the signal for half a period corresponding to a λ/4 = 327.5 nm displacement and find a local linear fit for the data (Ext. Data Fig. 6e). Knowing the displacement and the moving speed of the fiber, we can easily establish a relation between the intensity of the signal and the displacement, and calculate the vibrations amplitude around the average position given by the linear fit by looking at the histogram of the positions of the cold finger (Ext. Data Fig. 6f). We measured oscillations of the cold finger around the mean position of ± 83 nm. In the experiment performed at low temperature, the impact of the roughness of the sphere combined with the vibrations of the cryostat induce therefore a strong distance uncertainty in the sub-100 nm regime.

7. Characterization of the photovoltaic cells

The fabricated photovoltaic cells (see Methods) are shown in Ext. Data Fig. 7a. Because the cells are made of InSb, a very low bandgap energy material, they need to be cooled to operate properly[15]. In order to verify this requirement, we performed I-V curve measurements in the dark for different cell temperatures from 30 K to 250 K (Ext. Data Fig. 7b). In this configuration, thermal radiation illuminating the cell comes only from the ambient-temperature environment. Above ∼110 K, the I-V characteristics are linear. They do not correspond to a diode behavior anymore but are those of a passive resistive device. In this case the thermally-generated carrier concentration is high and the p-n junction effect does not exist anymore. When temperature decreases the exponential shape of the curve progressively appears. The reverse bias current rises while the forward bias current decreases. We observe in the semi-logarithmic scale (Ext. Data Fig. 7c) that for T ≤ 90 K, the open circuit voltage ($V_{OC}$) becomes positive and keeps increasing when the cell temperature is decreasing. The presence of a positive $V_{OC}$ means that the cell is generating power due to the 300 K ambient illumination. Thus the cell needs to be cooled to work properly. For our experiments, we chose a working temperature of 77 K as it corresponds to the commonly-used boiling point of liquid nitrogen. In addition, the illumination level provided by the emitter is high so the generated current is large enough to be detected at this temperature and does not require a lower temperature cooling to reduce the dark current. Note that in the main text (Fig. 2), it can be seen that the I-V characteristics in the dark exhibits a slightly-positive value for the short-circuit current. This is due to a slight offset of the electronics (2-3 nA), which was not subtracted in the results presented in the article.

The I-V characteristics indicate low series resistance in the devices. This is especially possible for micron-sized photovoltaic cells. However, the series resistance losses raise with current, which would become huge in large area photovoltaic cells. Smart designs of the front electrode have been recently discussed in this respect[57,58] and could help in upscaling.

8. Experimental results in the far field showing excellent agreements with the sphere-disc view factor

The experimental setup allows the measurement of the short-circuit current of the cell for emitter-cell distances up to 2 mm. In order to analyze the contribution of the far field to the photocurrent generation, we measured the short-circuit current ($I_{SC}$) of a cell having a 20 µm diameter active area with the emitter moving from the contact up to $d > 150$ µm, where the current starts to level off due to the low illumination. In this case, we used the *z*-piezo positioner in slip-stick mode with 80 nm steps. Then we compared the measured data to the evolution predicted by the analytic expression of the sphere-disc view factor given in Eq. (S4). Ext Data Fig. 8b (same data as that in the inset of Fig. 3b, different experiment than that reported in the other panels of Fig. 3) shows that the evolution of the short-circuit current matches well the prediction of the view factor from a few micrometers to more than 150 µm. This analysis shows that the far-field thermophotovoltaic conversion efficiency does not depend on distance. Below 2-3 µm, the measured data and the view factor prediction are not in agreement because the evanescent waves are contributing to the radiative heat transfer in addition to the propagative wave contribution predicted by the view factor (insets, Ext. Data Fig. 8b and Fig. 3b).

We also measured the evolution of $I_{SC}$ as a function of the lateral displacement of the emitter along the *x* or *y* axis, as it is the main parameter used to accurately position the emitter above the center of the active area of the cell (Ext. Data Fig. 8c). In this example the hot emitter was placed at $d \approx$ 10 µm from the cell surface and was moved laterally from approximately -120 µm to +120 µm relative to the center of the cell, while the short-circuit current was measured. As expected we observed a maximum for the current when the emitter is above the center (see optical-microscopy view from top in Ext. Data Fig. 8a), which first decreases rapidly and then more smoothly as the emitter goes away from the center. The measurements are shown in Ext. Data Fig. 8c., together with the theoretical model using the view factor, with a normalization applied at the maximum value. It can be noted that the measurements are slightly non-symmetrical. This issue is mainly due to the fact that the motion of the *x* and *y* positioners is not always perfectly smooth and linear, and can vary slightly over large displacements (larger than 100 µm). Most importantly, performing this kind of measurement along both *x* and *y* directions for a lower displacement range (approximately the size of the active area) provides a precise positioning of the emitter above the center of the cell. This is performed at vertical distances lower than 5 µm.

9. <u>Validation of the superposition principle</u>

For photovoltaic cells, the superposition principle tells that the measured photocurrent is equal to the sum of the current generated in dark conditions and the short-circuit current under illumination[59] in low-injection conditions[60]. So theory suggests that the shape of the I-V curve remains the same and is just shifted in $I_{SC}$ depending on illumination. This principle is very interesting experimentally because the entire I-V curve could be retrieved by measuring only the curve in the dark and then measuring $I_{SC}$ as a function of illumination. In order to verify the superposition principle, we started by measuring the I-V curve of a cell having a 20 µm active area diameter under dark conditions with a cooled radiative shield over the active area of the cell to block the ambient radiation coming from the environment. Then we used the emitter to provide different levels of illumination to the cell by either changing the emitter-cell distance or the emitter temperature (Ext. Data Fig. 9a). As expected we observed that when illumination increases the I-V curves are lowered into the photogeneration quadrant, corresponding to larger $I_{SC}$ and $V_{OC}$. It is worth noticing that the fill factors range from 0.69 to 0.75 at the highest illumination, which is remarkable for a cell with a low bandgap. Then we shifted each curves in current only so their $I_{SC}$ were all set equal to 0 (Ext. Data Fig. 9b). We observe that all the curves are well superimposed, so the superposition principle is valid for this kind of TPV cell. The fact that the superposition principle is valid is a proof that the temperature of the cell remains the same (77 K). Since the shape of the I-V curve strongly depends on temperature (see Ext. Data Fig. 7b,c), the curves would not be superimposed if the cell temperature was modified.

## 2. Supplementary Discussion

10. <u>Summary of the different studied configurations</u>

In the present work, we studied different parameters such as the emitter temperature, thickness of the cell substrate, diameter of the active area and p-doping level of the top layer. Ext. Data Tab. 1 sums up the main results obtained for each configuration, using either the superposition principle (*i.e* "Sup. pr.") or full measurements by scanning the whole voltage range (*i.e* "$\neq$ V."). The best enhancement factor of 5.9 and electrical power density of 7.5 kW.m$^{-2}$ were obtained with the sample having the largest p-doping concentration.

Although the melting point of InSb is 800 K, it was possible to make experiments with the emitter heated beyond this temperature (900 K) without degrading the cell at contact, most probably because the thermal contact resistance between the emitter and the cell limits the cell heating at contact. When the emitter is too hot, however, the cell can be locally heated at $T > 800$ K in the contact region. An attempt made with an emitter at 1200 K led to a degradation of the cell. As a consequence, we know that the threshold for the contact between the emitter and the cell is for an emitter temperature between 900 K and 1200 K. This issue does not prevent possible measurements with very hot emitters close to the cell, however the exact distance separating them cannot be measured easily with our method (see Ext. Data. Fig. 1c).

Finally, it must be reminded that the thermal conductance measurements require averaging over many curves. At room temperature, this is straightforward and about 100 curves allow obtaining a satisfying signal-to-noise ratio in a decent time (few hours). At low temperature, the permanent cooling of the cold finger, which does not reach quickly the stationary regime in regions far from the sample, induces a displacement, in particular laterally, of the sample that requires manual realignment of the sphere in front of the cell. This means that the time to perform experiments is much longer. As a result, averaging was performed on a smaller number of data than when done at room temperature and the near-field thermal conductance appears noisier. The long acquisition time is also the reason why the near-field efficiency was not always determined in the experiments reported in Tab. S1. It must also be noted that the estimation of the far-field contribution has some influence on the determination of the near-field efficiency, which can lead to some uncertainty. As a consequence, the near-field efficiencies provided in Tab. S1 should be considered with an uncertainty of ~15 %.

**Supplementary References**

**Corresponding author**

Rodolphe Vaillon (rodolphe.vaillon@ies.univ-montp2.fr).